# A low-power circuit for piezoelectric vibration control by synchronized switching on voltage sources

Hui Shen[a], Jinhao Qiu[a,*], Hongli Ji[a], Kongjun Zhu[a], Marco Balsi[b], Ivan Giorgio[c], Francesco Dell'Isola[d]

[a] The Key Laboratory of Education Ministry on Aircraft Structural Mechanics and Control, Nanjing University of Aeronautics & Astronautics, Yudao Street 29, 210016 Nanjing, China
[b] Dipartimento di Ingegneria Elettronica, Università di Roma Sapienza, via Eudossiana 18, 00184 Roma, Italy
[c] Dipartimento di Meccanica e Aeronautica, Università di Roma Sapienza, via Eudossiana 18, 00184 Roma, Italy
[d] Dipartimento di Ingegneria Strutturale e Geotecnica, Università di Roma Sapienza, via Eudossiana 18, 00184 Roma, Italy

**ABSTRACT**

In the paper, a vibration damping system powered by harvested energy with implementation of the so-called SSDV (synchronized switch damping on voltage source) technique is designed and investigated. In the semi-passive approach, the piezoelectric element is intermittently switched from open-circuit to specific impedance synchronously with the structural vibration. Due to this switching procedure, a phase difference appears between the strain induced by vibration and the resulting voltage, thus creating energy dissipation. By supplying the energy collected from the piezoelectric materials to the switching circuit, a new low-power device using the SSDV technique is proposed. Compared with the original self-powered SSDI (synchronized switch damping on inductor), such a device can significantly improve its performance of vibration control. Its effectiveness in the single-mode resonant damping of a composite beam is validated by the experimental results.

© 2010 Elsevier B.V. All rights reserved.

## 1. Introduction

Self-powered structural damping systems using piezoelectric elements have been intensively studied in the last few years. Such techniques make their application more popular, especially in aeronautics, where vibration control problems arise and the use of batteries or external power is restricted [1,2].

The simplest self-powered piezoelectric structural damping system is passive control, such as shunting piezoelectric elements with matched inductor or resistor [3]. However, this method has two disadvantages: the first one is the huge values of the inductor, which typically reaches hundreds of Henries in low-frequency domain. Synthetic impedance circuits overcome the problem of very large inductance values [4], but require an external power supply. And the second disadvantage is sensitive to environmental factors such as outside temperature, which cause drift in the structure's resonance frequencies. Once detuned, the shunt circuit loses its damping performance.

To overcome these drawbacks, while keeping simplicity of implementation and compactness, some switch shunting or semi-passive methods have been proposed [5–10]. However, they are generally implemented with either a digital signal processor or an analogue circuit with operational amplifiers, which need external energy. Particularly, the so-called synchronized switch damping on inductor (SSDI) technique has attracted more attention than others [11–17]. This method offers several advantages: it is insensitive to environmental changes due to its self-adaptive broadband behavior, it does not require a very large inductor for low frequencies, and multi-modal damping is achievable without complex circuits. Moreover, this technique only needs very low-power to operate the switch.

Up to now, several versions of self-powered switching circuits based on the SSDI technique have been developed [18–21]. However, it is noticed that a vibration damping system using only one piezoelectric patch and a 10 V voltage source, which is called SSDV (synchronized switching damping on voltage sources), appears almost as effective as another system using six patches with the SSDI approach [13]. Thus, it is of great interest to study the possibility of designing a "self-powered" system using SSDV technique.

In this paper, a vibration damping system powered by harvested energy with implementation of the SSDV technique is designed and investigated. A low-power version of the switching circuit for non-linear processing of voltage has been designed, and by combining the switching circuit with an energy harvesting subsystem, a vibration damping system using the SSDV technique without external power supply is established. It is shown that it also can deal with vibration control as effectively as the SSDI technique.

The paper is organized as follows. The second section summarizes the SSD technique and presents the theoretical analysis of increased power dissipated due to the SSDV technique. While the next section addresses the low-power circuit design. Operation of the circuit is demonstrated by the experiment in the fourth section. In addition, experimental results also confirmed the theoretical analysis of increased power dissipated due to the SSDV technique. For the sake of simplicity, two mechanical structures are used in the experiment: one is to supply energy to the circuit and the other is to test the effective of structural vibrating damping. Finally, the fifth section briefly concludes the paper by summarizing main results and exploring prospects for further work.

---

* Corresponding author. Tel.: +86 25 84891123; fax: +86 25 84891123.
E-mail address: qiu@nuaa.edu.cn (J. Qiu).

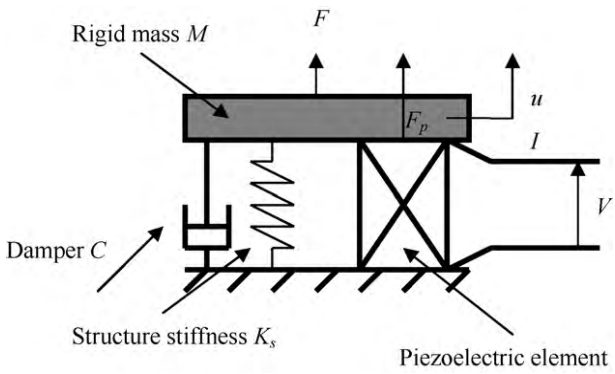

Fig. 1. Schematic representation of the electromechanical model.

## 2. Theoretical analysis of SSD technique

### 2.1. Modeling of the vibration system

As well known, if a structure vibrates near a resonance, an electromechanical model based on a spring-mass system with single degree of freedom gives a good description of the vibrating structure behavior (Fig. 1). The equation of motion of the spring-mass system can be expressed by:

$$M\ddot{u} + C\dot{u} + K_E u = \sum F_i \quad (1)$$

where $M$ represents the equivalent rigid mass and $C$ is the mechanical losses coefficient, $K_E$ is the equivalent stiffness of the electromechanical structure when piezoelectric elements are short-circuited, $u$ is the rigid mass displacement and $\sum F_i$ represents the sum of external forces applied to the equivalent rigid mass, including forces produced by piezoelectric elements, $F_p$, and force of the external excitation, $F$.

Piezoelectric elements bonded on the considered structure ensure the electromechanical coupling between the electrical system and the mechanical system, which is described by:

$$F_p = K_{PE} u + \alpha V \quad (2)$$

$$I = \alpha \dot{u} - C_0 \dot{V} \quad (3)$$

where $F_p$ is the electrically dependent part of the force applied by piezoelectric elements on the structure, $C_0$ is the blocked capacitance of piezoelectric elements, $K_{PE}$ is the stiffness of piezoelectric patches when it are short-circuited, and $\alpha$ is the force factor, $I$ is the outgoing current from piezoelectric elements.

Substitution of Eq. (2) into Eq. (1) gives:

$$F = M\ddot{u} + C\dot{u} + (K_S + K_{PE})u + \alpha V. \quad (4)$$

The following energy equation is obtained by multiplying both sides of Eq. (4) by the velocity and integrating over the time variable:

$$\int_0^t F\dot{u}dt = \int_0^t M\ddot{u}\dot{u}dt + \int_0^t (K_S + K_{PE})u\dot{u}dt + \int_0^t C\dot{u}^2 dt + \int_0^t \alpha V\dot{u}dt. \quad (5)$$

The provided energy is divided into kinetic energy, potential elastic energy, mechanical losses, and transferred energy. The transferred energy $E_S = \int \alpha V \dot{u} dt$ corresponds to the part of the mechanical energy which is converted into electrical energy. Maximization of this energy leads to minimization of the mechanical energy in the structure. The objective of all the SSD control approaches is to maximize this energy.

### 2.2. The principle of SSD technique

The principle of the SSD technique is to invert the voltage of the piezoelectric element (SSDI or SSDV technique) by briefly switching the shunt circuit consisting of the inductor $L$ (and an additional voltage source $V_S$ in the case of SSDV) to the closed state when the displacement $u$ reaches a maximum or minimum. In the closed state, the capacitance $C_0$ of the piezoelectric elements and the inductance $L$ constitute a resonance circuit. If the switch is closed just for half a period of the $L$–$C_0$ circuit ($t_i = \pi\sqrt{LC_0}$), the voltage on the PZT patch can be inverted as shown in Fig. 2. The voltage inversion time $t_i$ is far less than the vibration period.

The principle of the SSDV technique is shown in Fig. 3 [14]. Due to the additional voltage source $V_S$ in the inductive shunt circuit, the inverted voltage on the piezoelectric patch is boosted. For easy understanding of the SSDV technique, a typical time history of the voltage on the piezoelectric element and the structural deflection is illustrated in Fig. 4 [14].

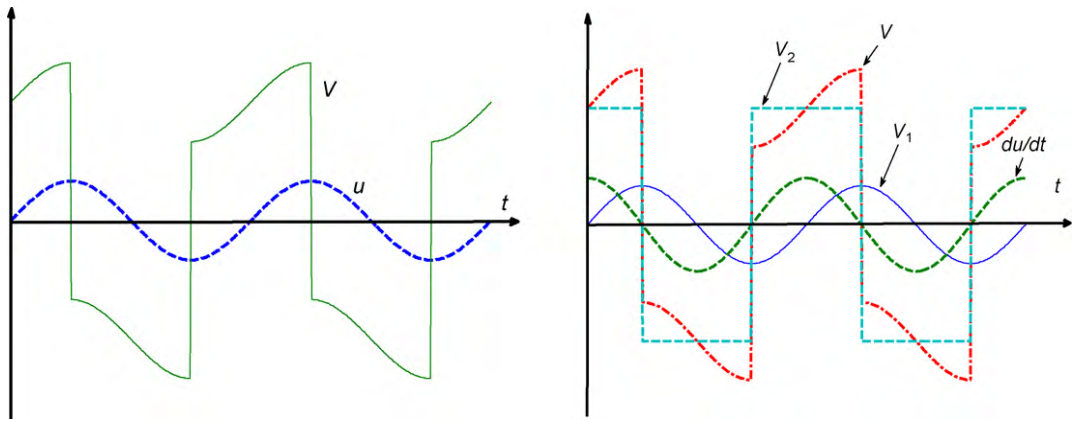

Fig. 2. Decomposition of the voltage $V(t)$ as the sum of two terms $V_1(t)$ and $V_2(t)$.

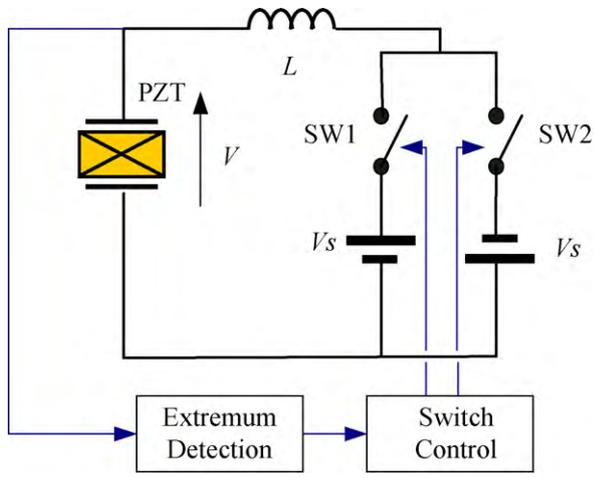

**Fig. 3.** The SSD device.

### 2.3. Energy analysis in SSD

It is assumed that the displacement remains purely sinusoidal during control. If the angular frequency of the driving force is $\omega$, the displacement can be expressed by:

$$u(t) = U_M \sin(\omega t + \varphi). \tag{6}$$

When the shunt circuit operates, the voltage of the piezoelectric element, $V$, can be expressed as the sum of two parts $V_1$ and $V_2$ as shown in Fig. 3 [15], where $V_1$ is an image of the displacement and $V_2$ is a rectangular waveform roughly proportional to $\text{sign}(\dot{u})$:

$$V = V_1 + V_2 = k_1 u + k_2 \text{sign}(\dot{u}). \tag{7}$$

It can be rewritten as:

$$V(t) = \frac{\alpha}{C_0}(u(t) + h(t)), \tag{8}$$

where $h(t)$ is the square function defined as [14]:

$$h = \frac{1+a_i}{1-a_i}\left(u_M + \frac{C_0}{\alpha}V_S\right)\text{sign}(\dot{u}), \tag{9}$$

in which $a_i \in [0, 1]$ is the voltage inversion coefficient. However, in the case of SSDI technique the square function $h(t)$ can be obtained by setting $V_S$ to zero. It is clearly shown that the SSDV technique increases the amplitude of voltage on the piezoelectric element. Thus, the transferred energy $E_S$ dissipated during a period $T$ by the SSDV process becomes:

$$E_S = \frac{\alpha^2}{C_0}\int_0^T h(t)du(t) + \frac{\alpha^2}{C_0}\int_0^T u(t)du(t) = \frac{\alpha^2}{C_0}\int_0^T h(t)du(t). \tag{10}$$

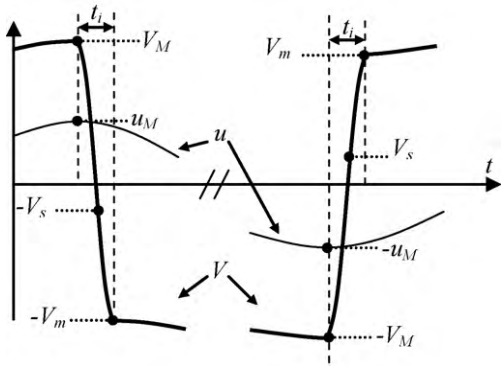

**Fig. 4.** The voltage on the PZT and the structural deflection.

Eq. (10) also stands for the SSDI technique. According to Ref. [11], the energy $E_S$ dissipated in SSDV process is:

$$E_S = 4\left(K_E k_{31}^2 U_M^2 + \alpha U_M V_S\right)\frac{1+a_i}{1-a_i}. \tag{11}$$

The dissipated energy of SSDI can easily be obtained by setting $V_S$ to zero. The damping is directly connected to the electromechanical coupling factor $k_{31}$ in the case of SSDI control. However, in the case of SSDV control the dissipated energy $E_S$ can be raised by increasing the voltage sources $V_S$, which will play a key role in the case of weakly coupled structure. Furthermore, the ratio of damping effect in the SSDV technique to the effect in SSDI control is shown as:

$$\frac{E_{SSDV}}{E_{SSDI}} = 1 + \alpha \frac{V_S}{K_E k_{31}^2 U_M}. \tag{12}$$

However, the voltage sources must deliver more energy to the system when the voltage $V_S$ is increased.

### 2.4. Power of the voltage sources in SSDV

As discussed above, voltage sources are inserted in the shunt circuit with implementation of the SSDV technique. Extra energy is delivered to the circuit as an electric current flow through the voltage sources. The average power of the delivered energy, $P_{V_S}$, can be expressed in the following form:

$$P_{V_S} = \frac{\int_0^T V_S I_S dt}{T}, \tag{13}$$

where $I_S$ is the electrical current through the voltage sources during voltage inversion. When the switch is briefly closed, the capacitance $C_0$ of piezoelectric elements and the inductance $L$ constitute a resonant circuit. The charge, $q_0$, on the piezoelectric element satisfies [22]:

$$L\ddot{q}_0 + r\dot{q}_0 + \frac{q_0}{C_0} = 0 \tag{14}$$

The current can be expressed as:

$$I_S = \dot{q}_0 = C_0 V_M \frac{\omega_0}{\sqrt{1-\xi_0^2}} e^{-\omega_0 \xi_0 t} \sin(\omega_0 \sqrt{1-\xi_0^2}\, t), \tag{15}$$

where $r$ is the resistor of the electrical loop, $V_M$ is the voltage across the piezoelectric element just before the switching event, $\omega_0$ and $\xi_0$ are the natural angular frequency and the damping coefficient of the resonant circuit, respectively, which are defined as:

$$\omega_0 = \sqrt{\frac{1}{LC_0}}, \quad \xi_0 = \frac{1}{2}r\sqrt{\frac{C_0}{L}}. \tag{16}$$

As introduced above the voltage inversion process lasts for half a period of the current oscillation in the shunt circuit in SSDI and SSDV controls. According to Eq. (15), the period of current oscillation is:

$$T_S = \frac{2\pi}{\omega_0\sqrt{1-\xi_0^2}}.$$

The energy delivered by the voltage source during each voltage inversion action is:

$$E_V = \int_0^{T_S/2} V_S I_S dt = V_S \int_0^{T_S/2} \dot{q}_0 dt = V_S C_0 (1+a_i) V_M. \tag{17}$$

Since the voltage on the piezoelectric element is inverted twice in each period of vibration, the power of energy delivery $P_{V_S}$ can be expressed as:

$$P_{V_S} = \frac{\int_0^T V_S I_S dt}{T} = \frac{2E_V}{T} = \frac{2V_S C_0 (1+a_i) V_M}{T}. \tag{18}$$

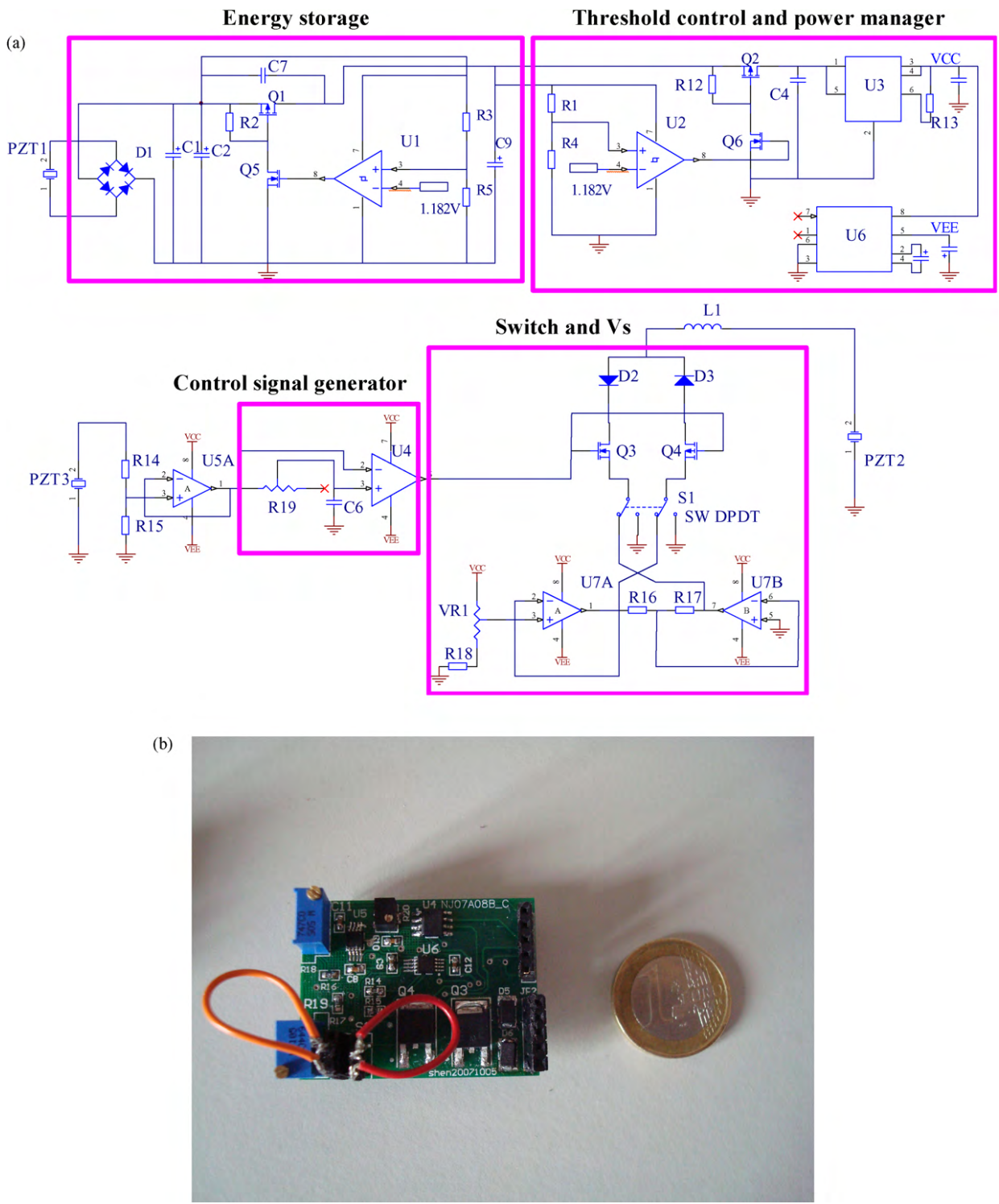

Fig. 5. Low-power vibration damping circuit: (a) schematic and (b) PCB prototype.

Hence it is clearly shown that the SSDV approach consumes energy while the SSDI technique does not consume energy (except for the switch energy).

## 3. Circuit design

The circuit mainly includes two parts: one is for energy harvesting and other is for the switch control circuit. The functions of the circuit for energy harvesting include voltage rectification, energy storage, threshold control and power manager. The switch control circuit includes the control signal generator, the switching circuit, and the voltage sources for SSDV. A schematic of the overall circuit is shown in Fig. 5(a) with the functions of different sub-circuits labeled in the same figure.

The energy is harvested from PZT1. The voltage signal from PZT1 is firstly rectified with a full-bridge rectifier, D1, and rectifier capac-itor, C1. The pulse-charging [23–25] circuit follows the rectifier circuit. The pulse-charger is comprised of an enhancement-mode p-channel MOSFET, Q1, an intermediate charging capacitor, C2, and a low-power comparator, U1. At first, the p-channel device, Q1, is kept close until the capacitor (C2)'s voltage reach 3.4 V. When the capacitor (C2)'s voltage reaches 3.4 V, the p-channel device Q1 is turned on by output of the low-power comparator U1. Thus the capacitor C2, 1 µF, is maintained at a relatively constant 3.4 V by comparator U1, which regulates the charging pulses to the storage capacitor C9, 1000 µF, through the p-channel device Q1.

The implementation of the threshold control is based on the charge state of the storage capacitor. The charge state of the storage capacitor is determined by the voltage of storage capacitor and measured by a comparator with hysteresis, U2. The comparator's negative reference voltage, along with the three resistors connected to the comparator's positive input, sets its hysteresis—that is, the values of high threshold voltage $V_{th(H)}$ and low threshold voltage $V_{th(L)}$. The threshold circuit's output turns the enhancement-mode p-channel MOSFET, Q2, "on" when the storage capacitor's voltage exceeds $V_{th(H)}$ and turn the p-channel device "off" when the storage capacitor's voltage drops below $V_{th(L)}$. That is, if the power of the energy harvesting system is not high, the control circuit can only work intermittently. When storage capacitor's voltage drops below $V_{th(L)}$, the threshold circuit stops supplying energy to the control circuit and the SSD technique is turned off. However, the pulse-charging of the capacitor C9 still continues. When the storage capacitor's voltage exceeds $V_{th(H)}$, the threshold circuit will start to supply energy again and operation of the SSD technique resumes. When the p-channel device (Q2) turns on, the current will flow from the storage capacitor C9 to a low-dropout linear regulator, U3, which supplies a positive regulated voltage, $V_{CC}$, to the control circuit. A switched-capacitor voltage converter, U6, which follows the linear regulators (U3), supplies a negative regulated voltage, $V_{EE}$, to the control circuit. Thus, all the chips in the two control sub-circuits (control signal generator and switch) are powered by the regulator, U3, and switched-capacitor voltage converter, U6.

The function of control signal generator is to detect the displacement extremum and generate the control signal. The first part of control circuit detects the displacement extremum from the open-circuit voltage of a piezoelectric element (PZT3). The instant of strain maximum is obtained by comparing two signals using a low-power comparator, U4. One signal is the voltage signal, and the other is the same signal with a little delay, which is obtained by a low pass filter consisting of a resistor and a capacitor. The low-power comparator generates the positive/negative voltage level according to the displacement maximum or minimum. Thus, electronic switch is tuned on/off by the control signal generated by the comparator.

As shown in Fig. 5, the electronic switch is composed of two MOSFET devices (NMOS Q3 and PMOS Q4), associated with two fast diodes, D2–D3. The manual switch (S1) can be used for switching between the SSDI technique and the SSDV technique. When the electronic switches, Q3 and Q4, are connected with the ground GND, the circuit operates in the SSDI mode. When the electronic switches, Q3 and Q4, are connected with two low-power operational amplifiers (U7A and U7B), the circuit works in the SSDV mode.

In the case of SSDV technique, two additional voltage sources $V_S$ are separately achieved by two operational amplifiers, U7A and U7B. Here the operational amplifier U7A works as a positive voltage source and the operational amplifier U7B work as a negative voltage source, whose output amplitude can be adjusted between 0 V and 2.5 V by tuning a variable resister VR1. Thus the two amplifiers (U7A and U7B), the electronic switch (Q3 and Q4), the piezoelement (PZT2) and the inductor (L1) constitute an electrical loop.

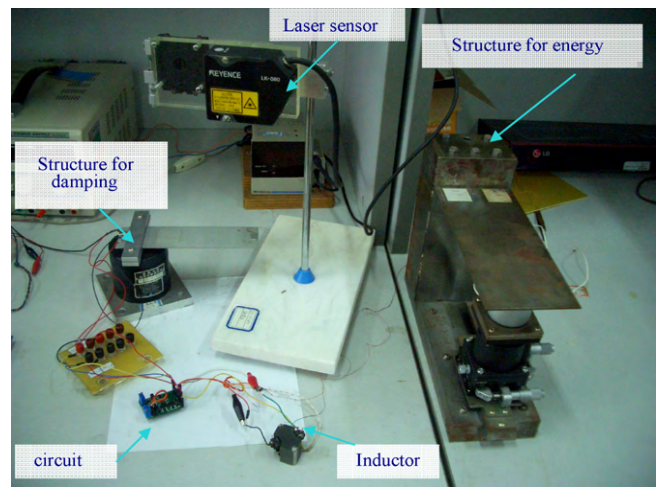

**Fig. 6.** Experiment equipment.

## 4. Experimental results

### 4.1. Experiment setup

In the experiment, there are two mechanical structures: one is to supply energy to the circuit and the other is to test the effective of structural vibrating damping. The first one, used as the piezoelectric energy source, is a cantilever steel beam clamped at one end to a rigid and heavy steel base. Two piezoelectric ceramic patches are bonded on the surface of beam, close to the clamped end. The external excitation force is provided by an electromagnet, driven by a signal generator and a power amplifier. The second structure, used for the damping test, is a cantilever beam clamped at one end to an electromagnetic shaker. Two piezoelectric ceramic patches are embedded in the beam, close to the clamped end. One can be used as a sensor (PZT3) and another as an actuator (PZT2). The poling direction is in the thickness direction of the beam. The free-end displacement of the cantilever beam is observed with Laser detector (KEYENCE LK-80). The experimental setup is represented in Fig. 6 and the main characteristics are given in Table 1.

### 4.2. Result and discussion

Experimental data were taken to validate the semi-passive system presented in this paper and to demonstrate the operation of circuit.

#### 4.2.1. The experiment of energy harvesting

At first, the characteristics of mechanical structure for supplying energy are identified. The displacement characteristics of the open-circuit and the short-circuit, measured in a frequency band around the first bending mode, are plotted in Fig. 7. The resonance frequencies and electromechanical coupling coefficient are deduced from the plot [3]. The parameters are presented in Table 2.

**Table 1**
Experimental setup characteristics.

| Parameter | Supply energy | Vibration damping |
|---|---|---|
| Beam material | Steel | GFRP |
| Beam size | 300 mm × 100 mm × 2 mm | 180 mm × 50 mm × 0.5 mm |
| 1st bending mode | 31.72 Hz | 25.84 Hz |
| Ceramic type | Z0.2T30 × 30S-SY1-C82 | Z0.2T30 × 30S-SY1-C82 |
| Patches size | 30 mm × 30 mm × 0.2 mm | 30 mm × 30 mm × 0.2 mm |
| Number of patches | 2 | 2 |

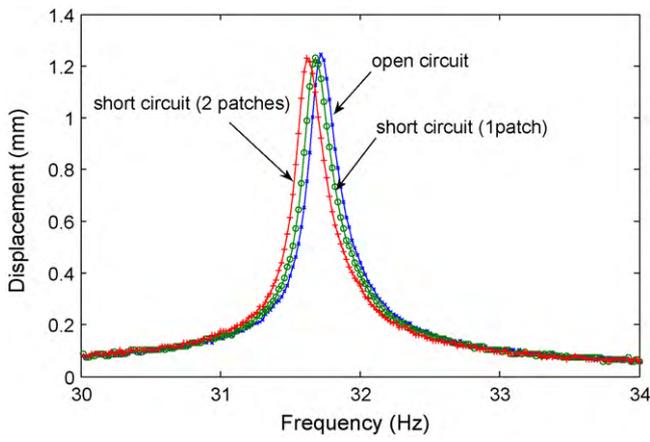

Fig. 7. Short-circuit and open-circuit characteristics.

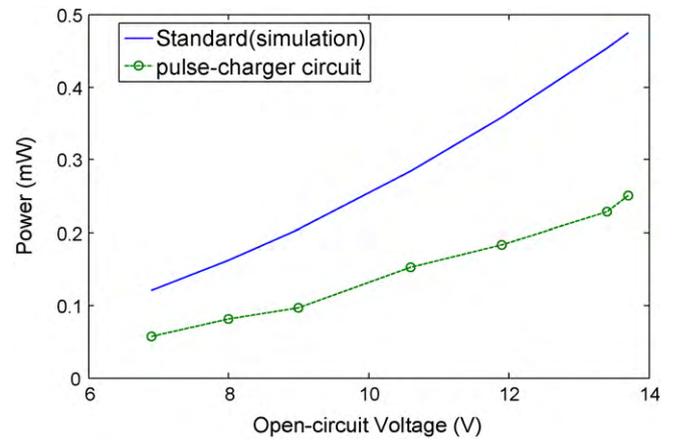

Fig. 8. Energy harvesting circuit performance.

The next experiment shows the performance of the energy harvesting circuit. Only one patch was connected with the pulse-charge converter in the experiment. For comparison purpose, it is calculated that available energy harvested by standard interface which includes a diode rectifier bridge and a filter capacitor [26]. The available power is determined by placing optimal resistor behind the standard interface, which is deduced by [24]:

$$P_{max} = \frac{\left(\frac{1}{2}V_{open}\right)^2}{R_{opt}} = \left(\frac{1}{2}V_{open}\right)^2 \frac{2C_0\omega}{\pi}, \quad (19)$$

where $V_{open}$ is the open-circuit voltage of piezoelement, $C_0$ is the clamped capacitance of the piezoelement, and $\omega$ is the resonant frequency of the structure. Additionally, Eq. (19) does not take into account the damping effect generated by the harvesting process. The mechanical excitation level of piezoelectric element is characterized by the piezoelectric element's open-circuit voltage $V_{open}$. As compared to the standard interface, the behavior of pulse-charger can be clearly seen in Fig. 8. Energy harvesting circuit losses, including the power consumption of the controller, can be calculated as the difference between the available power and that harvested with the converter. The experimental efficiency of the pulse-charger was between 45% and 55% in different excitation levels, as shown in Fig. 9. However, the power of energy harvested by the pulse-charger is not dependent on the load.

### 4.2.2. The experiment of structural vibration damping

The parameters of the simplified model of the structure in Fig. 1 are firstly identified. The short-circuit resonance frequency (1 patch), $f_0$, open-circuit resonance frequency, $f_1$, open-circuit damping coefficient, $\zeta$, the ratio of piezoelectric open-circuit voltage to the free-end displacement of the beam, $\gamma$, and the clamped capacitance of the piezoelectric element are measured experimentally and the other parameters of the simplified model are calculated from the following equations [13]:

$$\alpha = \gamma C_0, \quad K_E = \alpha\gamma\frac{f_0^2}{f_1^2 - f_0^2}, \quad M = \frac{K_E}{4\pi^2 f_0^2}, \quad C = 4\pi\xi M f_1. \quad (20)$$

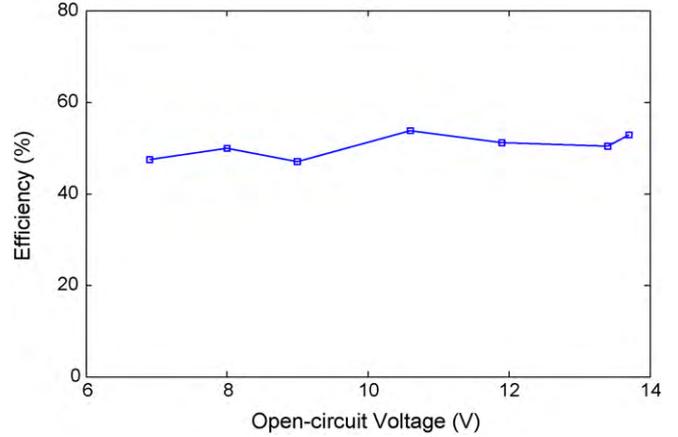

Fig. 9. Energy harvesting circuit efficiency.

The identified parameters are listed in Table 3.

For comparison, experiments were carried out in the following three different cases: (1) vibration control using the low-power vibration damping system in Fig. 6; (2) vibration control using the same system except that laser sensor with external power was used as a sensor instead of the piezoelectric element (PZT3); (3) vibration control using the standard SSDI technique implemented in a DSP environment based on a dSPACE board DS1103 [27].

**Table 2**
Experimental data.

| Parameter | Symbol | Value |
|---|---|---|
| Short-circuit resonance frequency (1 patch) | $f_{01}$ | 31.68 Hz |
| Short-circuit resonance frequency (2 patches) | $f_{02}$ | 31.62 Hz |
| Open-circuit resonance frequency | $f_{11}$ | 31.72 Hz |
| Electromechanical coupling coefficient (1 patch) | $k_{S1}$ | 0.0503 |
| Electromechanical coupling coefficient (2 patches) | $k_{S2}$ | 0.0796 |
| Clamped capacitance of the piezoelectric element of the beam for energy harvesting (1 patch) | $C_{01}$ | 84 nF |
| Clamped capacitance of the piezoelectric element of the beam for energy harvesting (2 patches) | $C_{02}$ | 168 nF |

**Table 3**
Experimental parameter.

| Parameter | Symbol | Value |
|---|---|---|
| Short-circuit resonance frequency (1 patch) | $f_0$ | 25.82 Hz |
| Open-circuit resonance frequency | $f_1$ | 25.84 Hz |
| Open-circuit damping coefficient | $\zeta$ | 0.0093 |
| Piezo open-circuit voltage to the beam free-end displacement ratio | $\gamma$ | 1.058824 kV/m |
| Clamped capacitance of the piezoelectric element | $C_0$ | 103 nF |
| Equivalent mass | $M$ | 2.8 g |
| Equivalent inherent damping | $C$ | 0.0085 N m$^{-1}$ s$^{-1}$ |
| Equivalent stiffness | $K_E$ | 74.5097 N m$^{-1}$ |
| Electromechanical coupling coefficient of the beam for vibration damping | $k_V$ | 0.039 |
| Voltage inversion coefficient | $a_i$ | 0.755 |

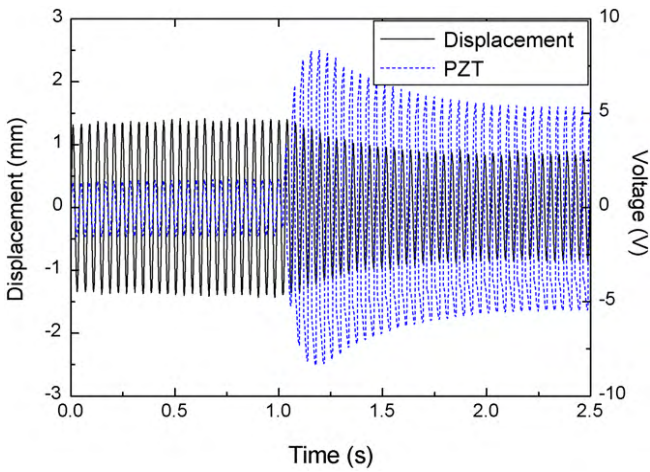

Fig. 10. The voltage of piezoelement using SSDV technique.

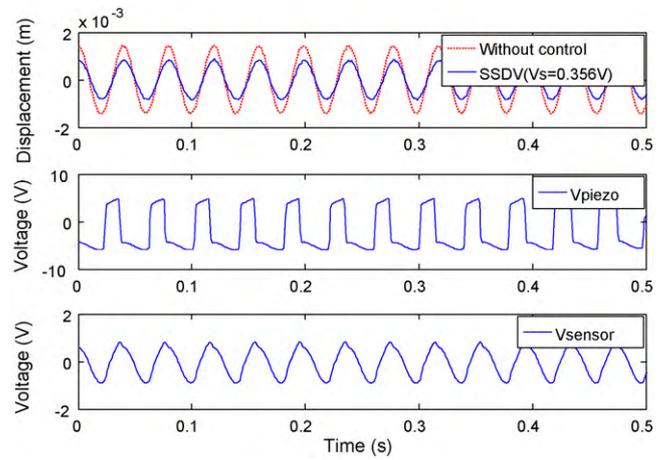

Fig. 12. Effectiveness of vibration damping using SSDV technique ($V_S$ = 0.365 V).

In the first case, the switching circuit starts to work only after the storage capacitor is charged to certain degree. The voltage on the piezoelectric element and the displacement of the free-end of the beam are shown in Fig. 10 when the beam is excited at the first resonance frequency. The voltage on the piezoelectric element is magnified and the displacement is reduced due to switching action of the circuit. For comparison purpose, the vibration of beam is also controlled by the SSDI technique. The zoom-in waveform of the voltage and displacement are shown in Fig. 11 for the SSDI technique and Fig. 12 for the SSDV technique. With the SSDI or SSDV technique, the voltage on the piezoelectric elements is inversed when the displacement $u$ reaches a maximum. However, it is clearly shown that the amplitude of voltage on the piezoelectric element with the SSDV technique is larger than the one using the SSDI technique. The amplitude of voltage on the piezoelectric element increases from 8.9 V by the SSDI technique to 10.6 V with the SSDV approach ($V_S$ = 0.356 V). Compared with a damping effect of 3.29 dB in SSDI, a vibration suppression of about 8.4 dB is achieved in the case of SSDV technique when the amplitude of the voltage source reaches 2.1 V. It agrees with the simulation result as shown in Fig. 13. The simulation was carried out using the Matlab/Simulink software. It also confirms the relation between the ratio of the damping effect and voltage source value $V_S$, which is defined in Eq. (12). However, it is observed that the difference between experiment results and simulation results become rather large when the voltage source value rises. The difference value increases from 1 dB ($V_S$ = 1 V) to 2.6 dB ($V_S$ = 2.1 V). Actually it is due to the sensor signal of the piezoelement. Compared with the sensor signal in the Figs. 11 and 12, the voltage of the piezoelement used as sensor does not give out an ideal signal when the voltage source value reaches 2.1 V, as shown in Fig. 14. It is mainly due to the influence of another piezoelement which is used as actuator. This kind of influence becomes strong when the amplitude of voltage on the piezoelectric element (used as actuator) is quite high. Thus, the distorted sensor signal leads to an inaccurate switching action. Even though the displacement $u$ does not reach a maximum, the switch has started to work.

On the other hand, if we use an ideal sensor signal, it can be predicted that the experimental damping value should be close to the simulation results. It is confirmed by the experimental result in the second case. As shown in Fig. 13, the experimental result remains close to the simulation result even in high value of voltage source. However, it is externally powered, not truly self-powered system.

In the third case, only a damping value of 4.6 dB is achieved with two piezoelements (PZT2 and PZT3) by the SSDI technique, as shown in Fig. 13. The damping effect using two piezoelectric ele-ments with the SSDI technique is almost the same as the low-power SSDV system using three piezoelectric patches and a 0.5 V voltage source (only using one piezoelectric material for harvesting purpose). Since the piezoelement used for sensor can be comprised by a quite small size of piezoelement in the low-power SSDV system (which can be omitted in comparison with the size of the energy harvesting patch or the piezoelectric insert used for damping), the area of piezoelectric patches in the low-power SSDV system can be almost the same as two piezoelectric elements with the SSDI technique. However, it can achieve better damping effect especially in the case of high voltage source value *Vs*, as shown in Fig. 13. On the other hand, it will require more energy in the case of high voltage source value *Vs* as predicted in Section 2. Therefore, a vibration damping system powered by harvested energy with implementation of the SSDV technique should take into account the relation between the higher voltage source and the more required energy, requiring a trade-off between better damping effect and the harvested energy.

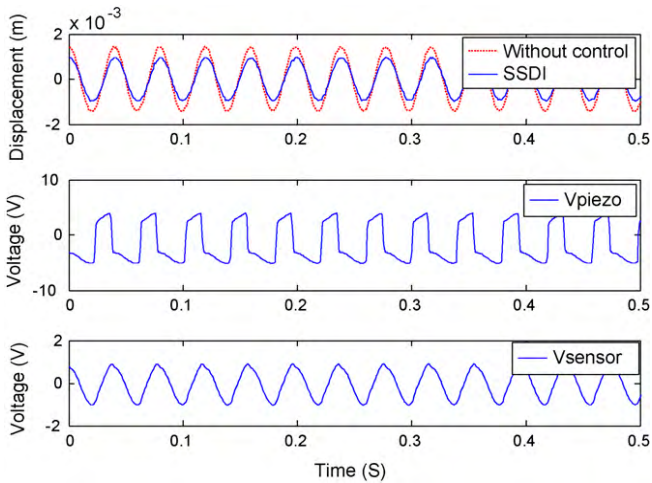

Fig. 11. Effectiveness of vibration damping using SSDI technique.

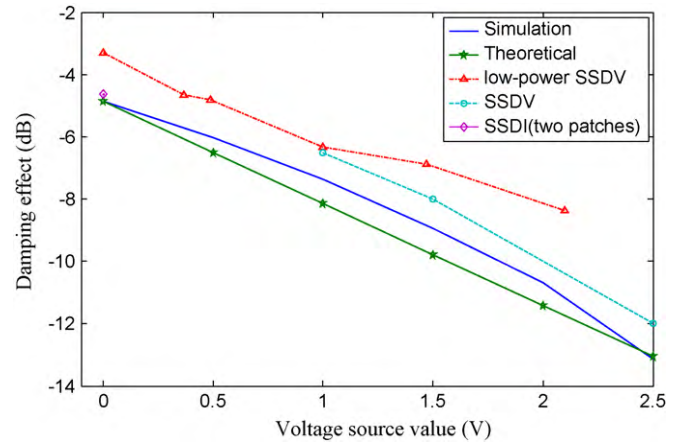

Fig. 13. Steady state damping at the resonance frequency.

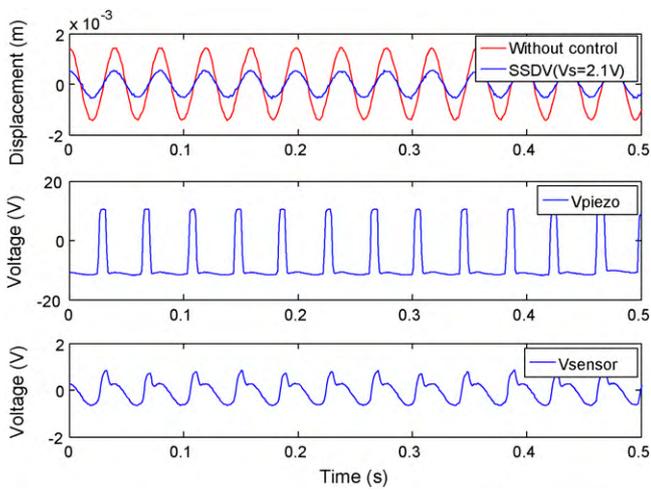

**Fig. 14.** Effectiveness of vibration damping using SSDV technique ($V_S$ = 2.1 V).

### 4.2.3. The power dissipation of the circuit

In the experiment, it is observed that the control circuit can only operate intermittently, as shown in Fig. 15. As stated in the section of circuit design, the threshold voltage has two levels: $V_{th(H)}$ and $V_{th(L)}$. When the voltage of the storage capacitor exceeds the high level $V_{th(H)}$, the linear regulator, U3, begins to output a voltage of 2.5 V and the control circuit starts to work. The electrical energy stored in the storage capacitor is dissipated by the control circuit. Since the rate of energy dissipation is higher than the power of harvested energy, the voltage of the storage capacitor decreases during vibration control. When the voltage of the storage capacitor drops below $V_{th(L)}$, the output of the linear regulator, U3, becomes zero and the control circuit stop working. At the same time, the voltage of storage capacitor increases again. The above charging and discharging processes of the storage capacitor are carried out repeatedly.

The time history of the output voltage, $V_{CC}$, of the regulator, U3, is shown in Fig. 15. The voltage of the piezoelectric element is also shown in the same figure. The system can be divided into two states: charging and operating. In the charging state, the storage capacitor is charged from two piezoelectric patches. In the operating state, the storage capacitor is discharged due to the power dissipation of the control circuit and also charged from the piezoelectric patches. The voltage of the piezoelectric element used as actuator is amplified and the damping effect is achieved in the operating state. If the durations of the charging and operating states are measured, the power dissipation, $P_{d(SSDV)}$, can be calculated from:

$$P_{d(SSDV)} = \frac{\left(V_{th(H)}^2 - V_{th(L)}^2\right)C_9}{2T_1} + \frac{\left(V_{th(H)}^2 - V_{th(L)}^2\right)C_9}{2T_2}, \quad (21)$$

where $T_1$ is the discharging time, $T_2$ is the charging time, and $C_9$ is the capacitance of the storage capacitor. In the experiments, the voltage levels, $V_{th(H)}$ and $V_{th(L)}$, were set to 3.1 V and 2.8 V, respectively. Thus, the total power dissipation for the control circuitry is about 691 μW in the SSDV mode ($V_S$ = 0.356 V). Here it is noted that the total power dissipation ($P_{d(SSDI)}$) in the case of SSDI technique is only about 322 μW [23].

Compared with the SSDI approach, higher power dissipation is mainly due to two parts: one is the increased power dissipation

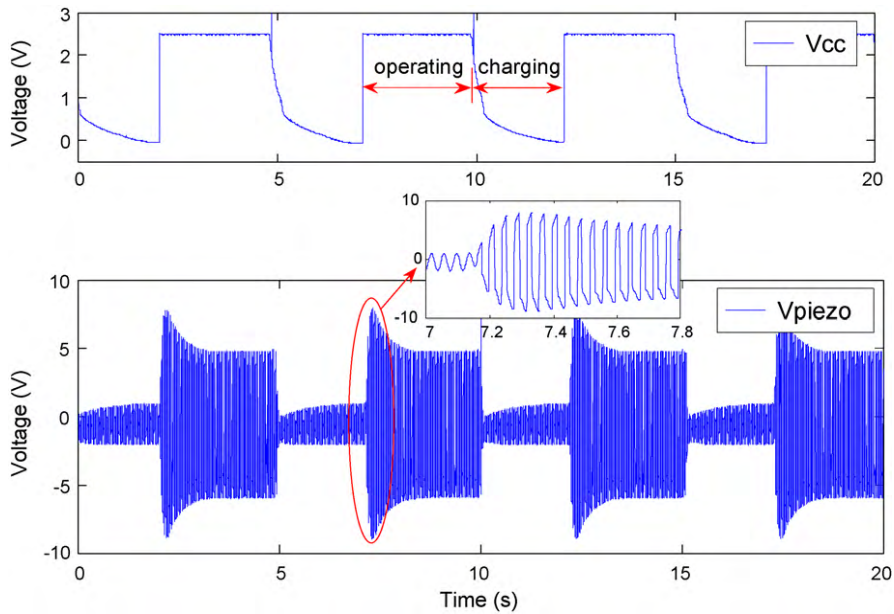

**Fig. 15.** The operation of the circuit ($V_S$ = 0.365 V).

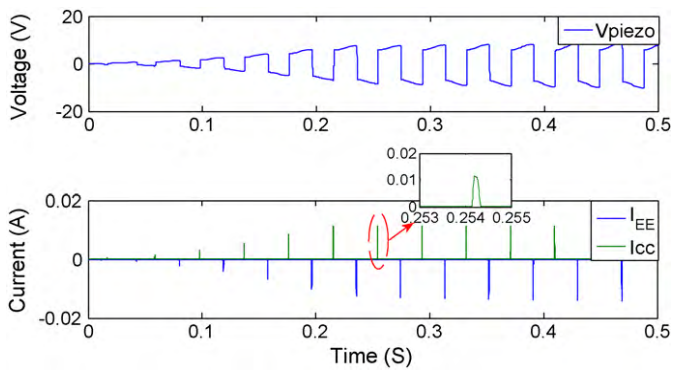

Fig. 16. The electrical loop in SSDV mode.

($P_{Vs}$) in the voltage sources in the SSDV technique and the other is the increased power dissipation ($P_{regulator}$) in positive/negative regulators (U3 and U6), which can be expressed as:

$$P_{d(SSDV)} = P_{d(SSDI)} + P_{Vs} + P_{regulator}. \quad (22)$$

The increased power dissipation ($P_{Vs}$) in the voltage sources occurs in two amplifiers U7A and U7B, as stated in the section of circuit design. It agrees with the simulation result as shown in Fig. 16. The simulation was carried out using the OrCAD software. In this case, the current $I_{CC}$ in the pin (positive power supply $V_{CC}$) of two amplifiers (U7A and U7B) immediately increases to the maximum and decreases rapidly to null during the switch turns on/off; the current $I_{EE}$ in the pin (negative power supply $V_{EE}$) of the same chip also has the same event during the switch turns on/off. Thus, the increased power dissipation ($P_{Vs}$) can be obtained from Eq. (18), as shown in Fig. 17. The increased power dissipation in the regulator, $P_{regulator}$, can be approximately estimated as:

$$P_{regulator} = \frac{V_{U3} \int_0^T (I_{CC} + I_{EE})dt}{T} + \frac{V_{U6} \int_0^T I_{EE}dt}{T}, \quad (23)$$

where $V_{U3}$ is the voltage difference between the input voltage and output voltage of the regulator (U3), $V_{U6}$ is the output voltage of regulators (U6). However, with the increase of power dissipation ($P_{d(SSDV)}$), a great improvement of damping effect is achieved, as shown in Fig. 18. At the highest value of voltage source, almost 2.1 V, the damping effect increases from 4.86 dB by the SSDI technique to 11.7 dB with the SSDV approach, over a factor of two improvements.

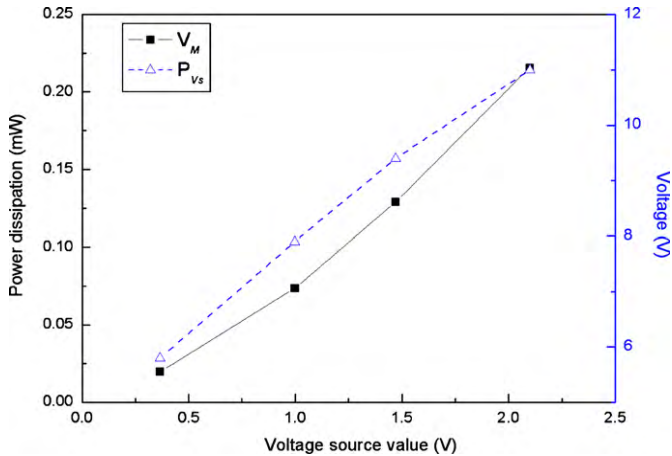

Fig. 17. The increased power dissipation in SSDV mode.

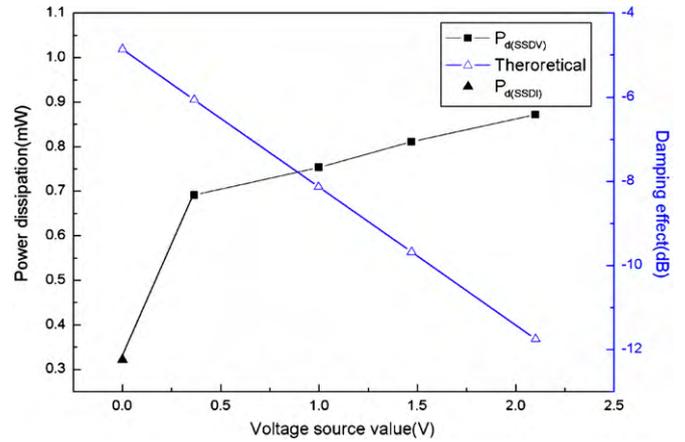

Fig. 18. The power dissipation of the circuit.

On the other hand, the energy harvested from two patches can be calculated from:

$$P_s = \frac{\left(V_{th(H)}^2 - V_{th(L)}^2\right) C_9}{2T_2}, \quad (24)$$

which is about 381 μW. Since the harvested energy $P_s$ is more than half of the power dissipation of the control circuit, it can be predicted that the control circuit will be able to work continually if four piezoelectric patches are used in energy harvesting. It is validated by the experiment. As shown in Fig. 19, the linear regulator outputs a constant voltage of 2.5 V and the control circuit operates continually when four piezoelectric patches are used for energy harvesting.

### 4.2.4. Implementation considerations

Actually, the low-powered circuit with implementation of the SSDV technique can also work in one mechanical structure which are bonded or inserted at least three piezoelectric elements. However, once the circuit work, the decrease of the vibration magnitude would lead to a decrease in terms of harvested power $P_s$ (in other words, suppressing too much vibration in one time – especially for the SSDV technique – would lead to less available energy for the energy harvesting). Due to the configuration of the circuit, the control circuit can still work, while the operating time is shortened. In addition, it also brings difficulty in valuing vibration damping effect or the power dissipation of the circuit. For the sake of simplicity, two mechanical structures are used in the experiment. Actually, there are also some examples of such situation in real application.

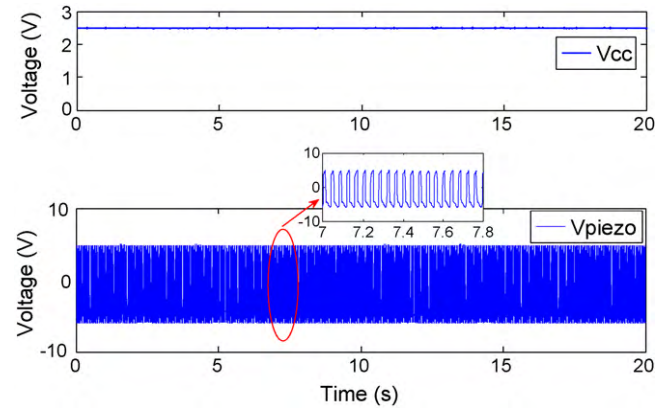

Fig. 19. The operation of the circuit powered by four patches ($V_S$ = 0.365 V).

For example, a piezoelectric microgenerator is placed on a vibration structure and supplies power to the low-power circuit which can achieve a great damping effect on other structure. Though in the experiment the structure used for energy harvesting is much larger than the controlled one, it is noticed that a quite small size of energy harvesting device can provide rather large energy according to Refs. [22,24].

## 5. Conclusion

This paper has designed a low-power circuit for piezoelectric vibration control by synchronized switching on voltage sources. Compared with the SSDI approach, a great improvement of damping effect is achieved by the SSDV technique. However, it has been theoretically and experimentally proved that the SSDV approach consumes more energy than the SSDI approach, which shows the difficulty in the implementation of a self-powered version of the SSDV technique. In other words, only enough energy (the delivered energy due to the voltage source) is supplied, it is possible to realize an autonomous vibration damping system by the SSDV technique. In the paper, it not only presents the theoretical and experimental analysis of increased power dissipated due to the SSDV technique, but also provides a possible solution which is under the framework of piezoelectric vibration damping and energy harvesting—that is, using energy harvesting technique to satisfy the need for more energy in comparison with the SSDI technique. Of course, the electronics on the control circuit should be optimized to decrease the power dissipation of the circuit in the future work, such as selecting more low-power electronic chips and components. In addition, more efficiency energy harvesting techniques should be under investigation in order to increase the power production.

## Acknowledgements

This research is supported by the National Natural Science Foundation of China (Grant Nos. 50775110 and 50830201) and Aeronautical Science Fund under Grant 20091552017. The authors would like to thank the Lab of intelligent material and structure (Università di Roma Sapienza), Fondazione Tullio Levi Civita and Comune di Cisterna di Latina for their support.

## References

[1] E. Garcia, K. Dosch, D.J. Inman, The application of smart structures to the vibration suppression problem, Journal of Intelligent Material Systems and Structures 3 (4) (1992) 659–667.
[2] V. Giurgiutiu, Review of smart-materials actuation solutions for aeroelastic and vibration control, Journal of Intelligent Material Systems and Structures 11 (7) (2000) 525–544.
[3] N.W. Hagood, A. von Flotow, Damping of structural vibrations with piezoelectric material and passive electrical networks, Journal of Sound and Vibration 146 (2) (1991) 243–268.
[4] A.J. Fleming, S. Behrens, S.O.R. Moheimani, Synthetic impedance for implementation of piezoelectric shunt-damping circuits, Electronic Letters 36 (18) (2000) 1525–1526.
[5] C.L. Davis, G.A. Lesieutre, An actively tuned solid-state vibration absorber using capacitive shunting of piezoelectric stiffness, Journal of Sound and Vibration 232 (2000) 601–617.
[6] W.W. Clark, Vibration control with state-switching piezoelectric material, Journal of Intelligent Material Systems and Structures 11 (4) (2000) 263–271.
[7] G.D. Larson, P.H. Rogers, W. Munk, State switched transducers: a new approach to high-power low frequency, underwater projectors, Journal of the Acoustical Society of America 103 (3) (1998) 1428–1441.
[8] K.A. Cunefare, State-switched absorber for vibration control of point-excited beams, Journal of Intelligent Material Systems and Structures 13 (2–3) (2002) 97–105.
[9] L.R. Corr, W.W. Clark, Comparison of low-frequency piezoelectric switching shunt techniques for structural damping, Smart Materials and Structures 11 (2002) 370–376.
[10] L.R. Corr, W.W. Clark, Energy dissipation analysis of piezoceramic semi-active vibration control, Journal of Intelligent Material System and Structure 12 (2001) 729–736.
[11] C. Richard, D. Guyomar, D. Audigier, G. Ching, Semi passive damping using continuous switching of a piezoelectric device, in: Proceedings of the SPIE International Symposium on Smart Structures and Materials: Passive Damping and Isolation, vol. 3672, San Diego, CA, 1999, pp. 104–111.
[12] C. Richard, D. Guyomar, D. Audigier, H. Bassaler, Enhanced semi-passive damping using continuous switching of a piezoelectric device on an inductor, in: Proceedings of the SPIE International Symposium on Smart Structures and Materials: Passive Damping and Isolation, vol. 3989, 2000, pp. 288–299.
[13] E. Lefeuvre, D. Guyomar, L. Petit, C. Richard, A. Badel, Semi-passive structural damping by synchronized switching on voltage sources, Journal of Intelligent Material System and Structure 17 (8–9) (2006) 653–660.
[14] A. Badel, G. Sebald, D. Guyomar, M. Lallart, E. Lefeuvre, C. Richard, J. Qiu, Piezoelectric vibration control by synchronized switching on adaptive voltage sources: towards wideband semi-active damping, The Journal of the Acoustical Society of America 119 (5) (2006) 2815–2825.
[15] L. Petit, E. Lefeuvre, C. Richard, D. Guyomar, A broadband semi-passive piezoelectric technique for structural damping, in: Proceedings of SPIE International Symposium on Smart Structure Materials: Damping and Isolation, vol. 5386, San Diego, CA, 2004, pp. 414–425.
[16] A. Faiz, D. Guyomar, L. Petit, C. Buttay, Wave transmission reduction by a piezoelectric semi-passive technique, Sensors and Actuators A 128 (2) (2006) 230–237.
[17] J. Qiu, G. Sebald, M. Yoshida, D. Guyomar, K. Yuse, Semi-passive noise isolation of a smart board with embedded piezoelectric elements, Electromagnetics Symposium Proceedings 17 (2005) 125–132.
[18] D. Niederberger, M. Morari, An autonomous shunt circuit for vibration damping, Smart Material and Structure 15 (2006) 359–364.
[19] M. Lallart, E. Lefeuvre, C. Richard, D. Guyomar, Self-powered circuit for broadband, multimodal piezoelectric vibration control, Sensors and Actuators A 143 (2007) 377–382.
[20] C. Richard, D. Guyomar, E. Lefeuvre, Self-powered Electronic Breaker with Automatic Switching by Detecting Maxima or Minima of Potential Difference Between its Power Electrodes, Patent # PCT/FR2005/003000, Publication number: WO/2007/063194 (2007).
[21] T. Yabu, I. Onoda, Non-power-supply semi-active vibration suppression with piezoelectric actuator, in: Proceedings of the JSASS/JSME Structures Conference, Vol. 47, 2005, pp. 48–50.
[22] M. Lallart, L. Garbuio, L. Petit, C. Richard, D. Guyomar, Double synchronized switch harvesting (DSSH): a new energy harvesting scheme for efficient energy extraction, IEEE Transactions on Ultrasonic, Ferroelectrics and Frequency Control 55 (10) (2008) 2119–2130.
[23] H. Shen, H. Ji, J. Qiu, K. Zhu, A semi-passive vibration damping system powered by harvested energy, International Journal of Applied Electromagnetics and Mechanics 31 (4) (2009) 219–233.
[24] G.A. Ottman, H.F. Hofmann, G.A. Lesieutre, Optimized piezoelectric energy harvesting circuit using step-down converter in discontinuous conduction mode, IEEE Transactions on Power Electronics 18 (2) (2003) 696–703.
[25] G.A. Lesieutre, G.A. Ottman, H.F. Hofmann, Damping as a result of piezoelectric energy harvesting, Journal of Sound and Vibration 269 (3–5) (2004) 991–1001.
[26] D. Guyomar, A. Badel, E. Lefeuvre, C. Richard, Toward energy harvesting using active materials and conversion improvement by nonlinear processing, IEEE Transactions on Ultrasonics, Ferroelectrics, and Frequency Control 52 (4) (2005) 584–595.
[27] H. Ji, J. Qiu, A. Badel, K. Zhu, Semi-active vibration control of a composite beam using an adaptive SSDV approach, Journal of Intelligent Material Systems and Structures 20 (4) (2009) 401–412.

## Biographies

**Hui Shen**, born in 1978, received B.S. degree in mechanical engineering from Shaanxi University of Science & Technology in 2000 and M.S. degree in mechanical engineering from Nanjing University of Aeronautics and Astronautics, in 2006. Currently he is working toward the Ph.D. degree in Aerospace Engineering, Nanjing University of Aeronautics and Astronautics, China. His current field of interest focuses on energy harvesting and vibration damping.

**Jinhao Qiu** received the B.S. and M.S. degrees in mechanical engineering from Nanjing University of Aeronautics and Astronautics, Nanjing, China, in 1983 and 1986, respectively. He received the Ph.D. degree in mechanical engineering from Tohoku University, Japan, in 1996. He was a research associate from 1986 to 1989 and lecturer from 1990 to 1991 at the Department of Mechanical Engineering, Nanjing University of Aeronautics and Astronautics. He was a faculty member at the Institute of Fluid Science, Tohoku University from 1992 to 2007, where he was a research associate from 1992 to 1998, an assistant professor 1998 to 2000, an associate professor from 2000 to 2004 and a professor from 2004 to 2007. Since March 2006, he has been a professor at the College of Aerospace Engineering, Nanjing University of Aeronautics and Astronautics. He has published more than 140 journal papers. His main research interest is smart materials and structural systems, including development of piezoelectric materials and devices, non-linear/hysteretic modeling of these materials, and their applications in vibration and noise control, energy harvesting, and active flow control in aerospace engineering.

**Hongli Ji**, born in 1983, received B.S. degree in mechanical engineering from Taiyuan Heavy Machinery Institute in 2004 and M.S. degree in mechanical engineering from Nanjing University of Aeronautics and Astronautics, in 2007. Currently she is working toward the PhD degree in Aerospace Engineering, Nanjing University of Aeronautics and Astronautics, China. Her current field of interest focuses on vibration damping, noise suppression and energy harvesting.

**Kongjun Zhu** was born in China, in 1971. He received his Ph.D. degree from Kochi University, Japan in 2005. In 2005–2007, he worked as an assistant professor in institute of fluid science at Tohoku University, Sendai city, Japan. He is presently a full-time university professor at Nanjing University of Aeronautics & Astronautics, Nanjing city, China. His research interests include hydrothermal synthesis of functional ceramics powder, preparation of biomaterials and fabrication of piezoelectric materials.

**Marco Balsi** received the M.Sc. and the Ph.D. in Electronic Engineering in 1991 and 1996, respectively, from "La Sapienza" University of Rome, Italy, where he is currently assistant professor, teaching circuit synthesis. He authored or co-authored about 90 refereed international publications. He is engaged in research on anti-personnel-mine detection systems, circuits for piezoelectromechanical structures, environmental and medical signal processing, artificial vision.

**Ivan Giorgio** was born in San Fele (Pz), Italy, in 1978. He received the B.E. (Mechanical Engineering) and Ph.D. degrees from The University of Rome "Sapienza", Rome Italy, in 2004, and 2008, respectively. He is a member of the Laboratory of Smart Structures and Materials in Cisterna di Latina (Lt). In 2009, he was a Postdoctoral Research Fellow at the University "Sapienza". His research involves passive, semi-active and self-sensing vibration control of piezoelectric laminates; acoustic noise control; and vibration control in robotic arms.

**Francesco Dell'Isola**, full professor in mechanics of solids, Università di Roma "La Sapienza", Dipartimento di Ingegneria Strutturale e Geotecnica, Laboratorio di Strutture e Materiali Intelligenti Cisterna di Latina, Master in Theoretical Physics and PhD in Mathematical Physics Università di Napoli "Federico II".